\documentclass[%
 aip,
 jmp,%
 amsmath,amssymb,
reprint,
]{revtex4-1}

\usepackage{graphicx}
\usepackage{dcolumn}
\usepackage{bm}

\begin{document}

\preprint{AIP/123-QED}

\title {Antiferromagnetic triangular Blume-Capel model with hard core
exclusions}

\author{A. Ibenskas}
\author{M. \v{S}im\.enas}
\author{E. E. Tornau}
\affiliation{Semiconductor Physics Institute, Center for Physical Sciences and
Technology, Go\v{s}tauto 11, LT-01108 Vilnius, Lithuania.}


\date{\today}

\begin{abstract}

Using Monte Carlo simulation we analyze phase transitions of two 
antiferromagnetic (AFM) triangular Blume-Capel (BC) models
with AFM interactions between third nearest neighbors. One model has hard core
exclusions between the nearest neighbor (1NN)
particles (3NN1 model) and the other - between 1NN and next-nearest-neighbor
particles (3NN12 model). Finite-size scaling
analysis reveals that in these models, as in the 1NN AFM BC model, the
transition from paramagnetic to long-range order (LRO)
AFM phase is either of the first-order or goes through intermediate phase which
might be attributed to Berezinskii-Kosterlitz
-Thouless (BKT) type. We demonstrate that properties of the low-temperature 
phase transition to the AFM phase of 1NN, 3NN1 and
3NN12 models are very similar in all interval of a normalized single-ion
anisotropy parameter, $\delta$, except for those values
of $\delta$, where the first order phase transitions occur. Due to different
entropy of the 3NN12 and 3NN1 models, their higher
temperature behavior is rather different from that of the 1NN model. Three phase
transitions are observed for 3NN12 model: 
(i) from paramagnetic phase to the phase with domains of the LRO AFM phase at
$T_c$ ; (ii) from this structure to diluted frustrated
BKT-type phase at $T_2$ (high-temperature limit of the critical line of the
BKT-type phase transitions) and (iii) from this frustrated
phase to the AFM LRO phase at $T_1$ (low-temperature limit of this line). For
the 3NN12 model $T_c>T_2>T_1$ at $0<\delta<1.15$ (range I),
$T_c=T_2>T_1$ at $1.15<\delta<1.3$ (range II) and $T_c=T_2=T_1$ at
$1.3<\delta<1.5$ (range III). For 3NN1 model $T_c=T_2>T_1$ at
$0<\delta<1.2$ (range II) and $T_c=T_2=T_1$ at $1.2<\delta<1.5$ (range III). In
range III there is only one first order phase transition.
In range II the transition at $T_c=T_2$ is of the first order, too. In range I
the transition at $T_c$ is either weak first-order or
second-order phase transition.

\end{abstract}

\maketitle

\section{Introduction}

The self-assembly of large triangular molecules attracts nowadays a great deal
of attention (see e.g. reviews~\cite{bartels,barth}).
The trimesic acid (TMA)~\cite{dmitriev,li1,ye,griessl,lackinger1,nath},
BTB~\cite{kampschulte,gutzler} and some other molecules
~\cite{theobald,silly,weber,li2,pawin} create patterns of different complexity
on solid-liquid interface or high-quality graphite and metal surfaces in ultra
vacuum conditions. The assemblies of such molecules might be different, but the,
so-called, honeycomb phase is the dominating low temperature pattern.

For description of the ordering of such (or similar) molecules the statistical
models of phase transitions might be used. The honeycomb phase might be
understood then as the low-temperature long-range order (LRO) phase on a
tripartite lattice in which the sites of each sublattice are occupied by
occupation variables $+1$, $-1$ and 0, respectively. For example, the ordering
of triangular TMA molecules might be described by the antiferromagnetic (AFM) nearest
neighbor (1NN) 3-state model~\cite{misiunas} which was originally
created to describe the ordering of lattice fluids and is sometimes called Bell-Lavis model~\cite{bell,fiore}. This model is similar to a better known triangular AFM lattice models: Blume-Capel (BC) model~\cite{blumecapel,mahan} with some neglected interactions, Blume-Emery-Griffiths model~\cite{beg} with anisotropic term~\cite{young,barbosa} and diluted triangular AFM Ising (TAFI) model~\cite{wannier}. 
The ordering of large molecules requires accounting for a finite size of the ordering objects, therefore some modifications related to hard core exclusions have to be introduced. To describe the ordering of TMA molecules into a series of flower phases, a model~\cite{ibenskas} with such exclusion at 1NN was proposed. In this model the initial lattice is rescaled, and the molecular interactions, which mimic the H-bonds, act between the molecules being on third nearest neighbor (3NN) sites, while the exclusion mimic the hard core infinite repulsive interaction occurring due to finite size of the molecules. 

At least two important questions arise when triangular AFM and other lattice models are used (see e.g.~\cite{weber,silly,petrauskas,fortuna,simenas1,simenas2}) to descibe the molecular ordering: to what extent the standard models might be rescaled and what effect the exclusion brings in comparison with classical (i.e. usually 1NN) statistical models. Intuitively, it is clear that the rescaling to the nNN models (with n$>1$) changes the entropy of
the system repressing the ordered phases and decreasing the phase transition
temperature if more sites for molecular diffusion occur. On the other hand, the
exclusion, which necessarily comes due to large size of the molecules, decreases
the number of sites for diffusion and, while promoting the ordered phases,
increases the transition temperature. 

Here we try to answer these questions using a triangular AFM BC model with
exclusions as an example. This model is tightly related to TAFI model which was
extensively studied~\cite{wannier} due to its frustrated phase and large entropy
at $T=0$. The TAFI model with magnetic field (chemical potential) gives rise to
ordered 3-sublattice structure, denoted as $\sqrt3\times\sqrt3$, in which
magnetizations (densities) of two sublattices are mutually equal but different
from that of the third~\cite{schick}. When diluted by the vacancies, which are
not fixed, but evolve together with the spins (the, so-called, annealed
vacancies), the TAFI model also allows for the 3-sublattice LRO structure.
Simple substitution of occupation variables~\cite{salinas} transforms a diluted
TAFI model in a field to Ising spin-1 or BC model~\cite{blumecapel}. 
The TAFI model also can be mapped into 6-state AFM clock model~\cite{miyashita}.
As shown by Cardy~\cite{cardy}, the 6-state clock model can exhibit  either a
first-order transition, two Berezinskii-Kosterlitz-Thouless (BKT) type
transitions or successive Ising, three-state Potts, or Ashkin-Teller-like
transitions. With decrease of temperature the 6-state AFM clock model on
triangular lattice gives a succession of two very close phase transitions which
are attributed to Ising (chiral) and BKT-type respectively~\cite{noh,surungan}.

The earliest study~\cite{mahan} of triangular AFM BC model performed by
renormalization group methods demonstrated that the AFM LRO phase, which has the
sites of its three sublattices occupied by variables 1, -1 and 0 respectively,
can be obtained on a phase diagram of this model. This phase occurs when the
ratio of a single ion anisotropy parameter normalized to  antiferromagnetic
coupling, $\delta=\Delta/|J|$, is within limits 0 and 3/2.  When $\delta>3/2$
the gas (disordered) phase prevails. It was shown in Ref.~\cite{mahan} that the
phase transitions to the LRO phase are of the second order for all $\delta>0$,
except at the very limit of the LRO phase, $\delta\rightarrow3/2$, where the
first order phase transition was found. Overstepping other limit of the LRO
phase, i. e. at $\delta<0$, the frustrated phase typical to the TAFI model
occurs. It should be noted that treating the model spin variables as particle
variables and using the lattice-gas rather than the magnetic formalism, the
single ion 
anisotropy parameter $\delta$ might be understood as a chemical potential. Then
decrease of $\delta$ is associated with increase of particle concentration at
expense of vacancies and transition from the 3-state AFM BC model to the 2-state
TAFI model (no vacancies) at $\delta=0$. 

Recently, a consistent study of phase transitions of this model was performed
using Monte Carlo (MC) calculations~\cite{zukovic}. It was clearly shown that
the phase transition from the paramagnetic to the AFM LRO phase is mediated by
the BKT-type phase in all interval of $\delta>0$ values, where the LRO AFM phase
exists, except for $\delta\gtrsim 1.47$, where the first order phase transition
between paramagnetic and LRO AFM phases is found.

In this paper we study the AFM BC model with exclusions. The problem is solved
assuming the AFM interaction of spins residing on 3NN sites. Two models with
hard core exclusions are considered: the 3NN model with exclusions at the 1NN
sites (3NN1) and the 3NN model with 1NN and 2NN exclusions (3NN12) (see Fig.
1). 
It should be noted that 3NN AFM BC model without exclusions is not studied here,
because it gives entirely different type of LRO AFM phase as its ground state
structure.

The obtained results for 3NN1 and 3NN12 models are compared with the results of
the 1NN model. We study how the exclusion affects the type of phase transition,
critical line of the BKT points and phase diagram.  
We demonstrate that in both 3NN models with exclusion the BKT-type phase
transitions obtained in 1NN model survive. Nevertheless, the phase transitions
in the 3NN1 and 3NN12 models are similar to those of the 1NN model only at lower
temperature, where the transition from the BKT-type phase to the LRO AFM phase
is found, and at those values of a single-ion anisotropy parameter which are not
close to the gas phase limit. At higher temperature and close to this limit the
properties are rather different. The exclusions decrease the high-temperature
point of the BKT-type phase transition and might stimulate the occurrence of a
phase transition from paramagnetic phase to the phase with domains of the LRO
AFM phase. It is demonstrated in the last section of the paper that the higher
temperature part of the phase diagrams of both 3NN models is very different from
that of the 1NN model. 
\begin{figure}
\includegraphics[width=0.9\columnwidth]{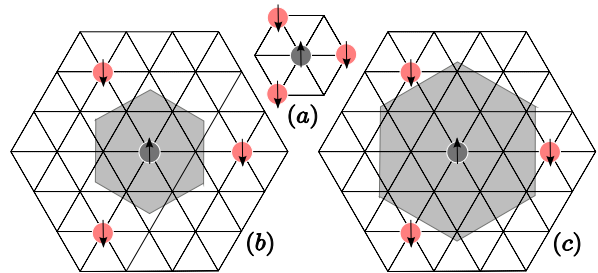} 
\caption{(color online) Particle (spin) arrangement in the LRO AFM phase on
triangular lattice for (a) 1NN, (b) 3NN1 and (c) 3NN12 models. Gray regions
schematically mark the limits of interaction exclusion (infinite repulsion) for the central spin.}  
\label{fig1.eps}
\end{figure} 
\

\section{Model and details of simulation}

The model Hamiltonian has the form

\begin{equation}
{\cal H}= -J\sum_{i,j}s_is_j+\Delta\sum_{i}s_i^2,
\end{equation}

where $s_i=\pm1,0$ is the spin variable on the triangular lattice site $i$, $J$
is the antiferromagnetic ($J<0$) interaction parameter acting between the
particles at 3NN sites, and $\Delta$ is a single-ion anisotropy parameter. Here
we regard the introduced variables as describing the magnetic particles in the
diluted lattice-gas model rather than the spin projections. Therefore in (1) we
write $\Delta$ with plus sign and treat this parameter as a chemical potential,
i.e. the total concentration of $\pm1$ particles increases (decreases) with
decrease (increase) of $\Delta$. Consequently, in the 3NN12 model the
interactions between particles separated by 1NN and 2NN distances are forbidden
by taking infinite repulsion of particles at these sites. In the 3NN1 model the
interactions between particles in the 1NN sites are forbidden in the same way.
Further, the temperature and single-ion anisotropy parameter are both normalized
to $|J|$: $k_BT/|J|$ and $\delta=\Delta/|J|$. 

Since cluster algorithms for frustrated systems are known to be
ineffective~\cite{coddington}, we performed the simulation of phase transition
properties using local update (single-flip) Metropolis
algorithm and Glauber dynamics. In the beginning the sites of a
triangular lattice were randomly populated by particles in states +1, −1 and 0,
and the initial energy $E_i$ of a randomly chosen molecule was calculated. Then the
initial state of that molecule was changed (with equal probability) to one of
two remaining
states, and the final energy $E_f$ was calculated. The new state was accepted,
if the
energy decreased after the change of state, or accepted with the probability
$\sim\exp[-(E_f-E_i)/k_BT]$,
if increased. Thus, the calculations were performed with fixed
chemical potential, while the concentration of particles in non-zero state, $c=\sum_is_i^2/L^2$, was allowed to vary.

For thermal averaging MC calculations and finite size scaling (FSS) of both 3NN
models with exclusions we used the triangular lattices of sizes $L\times L$ with
$L$ from 96 up to 216. For calculations of the 1NN model, which we performed to
compare the results, the lattice sizes $L=48$, 72, 96, 120 were used (for 1NN
model $J$ in (1) is acting between the particles on the 1NN sites). We used
periodic boundary conditions and $(0.2-1)\times10^6$ MC steps (MCS) for
thermalization. Further, we collected averages of 10$^7$ MCS for the 3NN models
and 10$^6$-10$^7$ MCS for the 1NN model. Our simulations were performed starting
from higher temperature and using random initial particle configuration. Then
the temperature was gradually decreased in small steps with simulations at new
temperature starting from the final configuration of the previous temperature.

Since phase transition parameters in this system were often characterized by
abruptness of their thermodynamic parameters and first order phase transitions,
in particular, we performed also energy histogram calculations using reweighting
techniques~\cite{reweight}. For these calculations we used slightly larger
lattice sizes ($L=120-270$) than for the thermal averaging. In some cases we
used a very large lattice size, $L=360$ and 399. Our simulations of thermodynamic
parameters (energy derivatives) often proceeded as follows: the phase transition
point was located by thermal averaging and then its slight correction was
performed by reweighting calculations.

We also performed the analysis of the autocorrelation time of energy at $T_c$
at $\delta=0.7$ for both 3NN models. The integrated autocorrelation
time  for the 3NN12 model ranged from $\tau\sim10^3$ MCS for $L=120$ to
$\tau\sim10^5$ MCS for $L=399$. For the 3NN1 model this time is around one-two
orders of magnitude higher. 

For studies of phase transitions we used the AFM order parameter. It should be
noted, that low temperature AFM phase of the 1NN model is stabilized when each
sublattice of the tripartite lattice is occupied by $+1$, $-1$ and 0 variables,
respectively. The distance between so-occupied sites is one lattice constant of
a triangular lattice, $a$. The stoichiometric particle concentration (coverage
of sites occupied by the $\pm1$ particles) in the 1NN model is $c_s=2/3$. Low
temperature AFM phase of both 3NN models
has 12 sublattices, only two of which are occupied by the $+1$ and $-1$
particles, respectively, and all other sublattices are empty. Therefore the
distance between  $+1$ and $-1$ particles  is $2a$ and $c_s=1/6$ in the AFM
phase of both 3NN models. As an order parameter, we use the staggered
magnetization, a slightly reworked version of the one suggested for the 1NN
model~\cite{zukovic}. It is the average difference of maximally and minimally
occupied sublattices. For 3NN model we had to account for occupancy of 12
sublattices, and therefore the staggered magnetization has the form

\begin{eqnarray}
m_s=\langle M_s\rangle/L^2 = \nonumber\\
6\Big\langle\mathrm{max}\Big(\sum_{i1\in\mathrm{sub1}}s_{i1},
\sum_{i2\in\mathrm{sub2}}s_{i2},...,\sum_{i12\in\mathrm{sub12}}s_{i12}\Big) - \nonumber\\
\mathrm{min}\Big(\sum_{i1\in\mathrm{sub1}}s_i,\sum_{i2\in\mathrm{sub2}}s_{i2},
...,\sum_{i12\in\mathrm{sub12}}s_{i12}\Big)\Big\rangle/L^2.
\end{eqnarray}

Here $i1...i12$ denote sites belonging to each sublattice, and the factor 6 is
needed to compensate for the stoichiometric concentration of the AFM phase in
the 3NN models. We calculate also temperature dependences of the specific heat
$C_v=(\langle{\cal H}^2\rangle-\langle{\cal H}\rangle^2)/L^2k_BT^2$,
susceptibility $\chi=(\langle M_s^2\rangle-\langle M_s\rangle^2)/L^2k_BT$, 
logarithmic derivatives of $\langle M_s\rangle$ and $\langle M_s^2\rangle$ 

\begin{eqnarray}
D_{1s}=\frac{\partial\ln\langle M_s\rangle}{\partial\beta}=\frac{\langle
M_s{\cal H}\rangle}{M_s}-\langle {\cal H}\rangle\\
D_{2s}=\frac{\partial\ln\langle M_s^2\rangle}{\partial\beta}=\frac{\langle
M_s^2{\cal H}\rangle}{M_s^2} -\langle {\cal H}\rangle\nonumber          
\end{eqnarray}

and Binder order parameter and energy cumulants, $U_B^{m}=1-\langle
M_s^4\rangle/3\langle M_s^2\rangle^2$ and $U_B^E=1-\langle {\cal
H}^4\rangle/3\langle {\cal H}^2\rangle^2$, respectively. The functions $D_{1s}$
and $D_{2s}$ were introduced in Ref.~\cite{ferrenberg}. They were
shown~\cite{zukovic} to be useful for a finite-size scaling of the 1NN AFM BC
model. At the second order phase transition point $T_c$ the maximum of specific
heat and susceptibility scale as $C_v\sim L^{\alpha/\nu}$ and $\chi\sim
L^{\gamma/\nu}$, respectively, while minimum of $D_{1s}$ and $D_{2s}$ - as $\sim
L^{1/\nu}$. Here $\alpha$, $\beta$ and $\nu$ are critical exponents of specific
heat, susceptibility and correlation length, respectively. At the first order
phase transition at $T_c$ the extrema of all these functions scale as $\sim
L^{d}$~\cite{challa}, where $d$ is dimensionality of the system. 

In a following section we present the values of critical exponent ratios $\alpha/\nu$ and $1/\nu$ at the phase transition point from the paramagnetic phase, $T_c$. 
The ratio $\alpha/\nu$ is obtained either by combined thermal averaging and reweighted histogram calculation of specific heat maximum at $T_c$ or by scaling these values close to $T_c$ and using the formula $C_v-C_0\sim L^{\alpha/\nu}f(tL^{1/\nu})$ (here $t=|T_c-T|/T_c$ and background is assumed to be $C_0=0$). The latter formula also gives the value of $1/\nu$ which we alternatively obtain as the average of scaling of parameters $D_{1s}$ and $D_{2s}$. 

In a case of BKT-type phase transitions correlation length diverges as
$\xi=\xi_0\exp\{a[(T_{\mathrm{BKT}}-T)/T_{\mathrm{BKT}}]^{-1/2}\}$ 
and spin-correlation function  decays as $\langle s_is_j\rangle\sim
r_{ij}^{-\eta}$, where $\eta$ is the critical exponent of the correlation
function~\cite{KT}. The order parameter at the BKT-type of phase transition
point scales as $m_s(L)\sim L^{-\eta/2}$. The exponent $\eta$ might also be
obtained from a part of susceptibility $\chi'= \langle M_s^2\rangle/L^2k_BT\sim
L^{2-\eta}$~\cite{challa2}. To obtain accurate values of the BKT-type phase
transitions of the AFM BC 1NN model at $T_1$ and $T_2$, the FSS of parameters
$m_s$ and $\chi'$ was performed~\cite{zukovic}. The following relations were
used

\begin{eqnarray}
m_sL^b=f_1\Big\{L^{-1}\exp\Big[a\Big(\frac{T_1-T}{T_1}\Big)^{-1/2}\Big]\Big\},
\hspace{0.2cm} T<T_1 \\\nonumber
\chi'L^c=f_2\Big\{L^{-1}\exp\Big[a\Big(\frac{T-T_2}{T_2}\Big)^{-1/2}\Big]\Big\},
\hspace{0.2cm} T>T_2
\end{eqnarray}

where $b=\eta/2$ and $c=2-\eta$ and $T_1$ and $T_2$ are lower-temperature AFM
LRO phase $-$ frustrated (BKT-type) phase and higher-temperature frustrated
phase $-$ paramagnetic phase transition temperatures, respectively.

\section{Results of simulation}

Both 3NN12 and 3NN1 models, as well as the 1NN model, have the low-temperature
AFM phase at values of the single-ion anisotropy parameter $\delta$ in between 0
and 3/2. The behaviour and properties of phase transitions are different for
different values of $\delta$. For the 3NN12 model we found three important
ranges of $\delta$ values. The range I corresponds to the situation when there
are three consecutive phase transitions: two of them are of the BKT-type ($T_1$
and $T_2$), and they frame the critical line of the BKT-type phase transitions,
and the third is a high-temperature phase transition at $T_c$. In range II there
remains the critical line of the BKT-type phase transition points, and $T_1$ as
its low-temperature point, but $T_2=T_c$. This range is found for both 3NN12 and
3NN1 models. In range III there is just one first order phase transition at
$T_1=T_2=T_c$. This range is found for both 3NN12 and 3NN1 models (as well as
for the 1NN model at $\delta>1.47$; in between 0 and 1.47 the 1NN model 
demonstrates just two BKT-type phase transitions at $T_1$ and
$T_2$~\cite{zukovic}).
 
\begin{figure}
\includegraphics[width=\columnwidth]{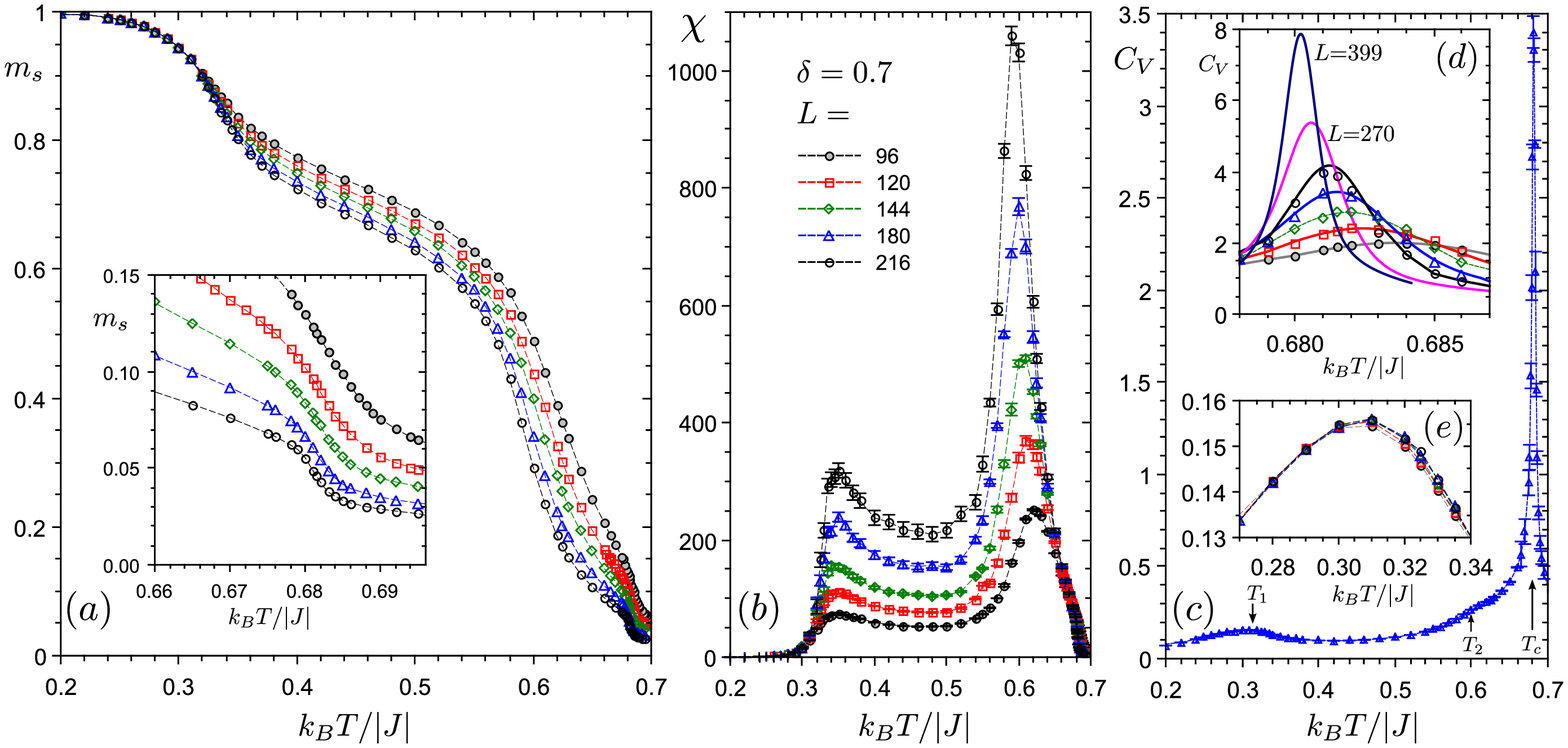}
\caption{(Color online) Temperature dependence of (a) staggered magnetization
and (b) susceptibility of the 3NN12 model at $\delta=0.7$ for different values
of $L$. Inset in (a): magnified behavior of $m_s(T)$ at $T_c$. Errors in (a) do
not exceed symbol size. (c) Temperature dependence of specific heat of the 3NN12
model at $\delta=0.7$ and $L=180$. Insets in (c) show $C_v(T)$ dependence around
$T_c$ (d) and $T_1$ (e) for different values of $L$. The symbols and solid lines
in (d) denote the results of thermal averaging and reweighting, respectively.} 
\label{fig2.eps}
\end{figure} 

\subsection{3NN12 model at $\delta=0.7$ (range I)}

The attraction of different particles at the 3NN sites and exclusion rules
imposed on 1NN and 2NN neighbors makes the 3NN12 model similar (just scaled-out)
version of the 1NN model. This is indeed the case at low temperature. The
situation at higher temperature is rather different.

\begin{figure}[] 
\includegraphics[width=\columnwidth]{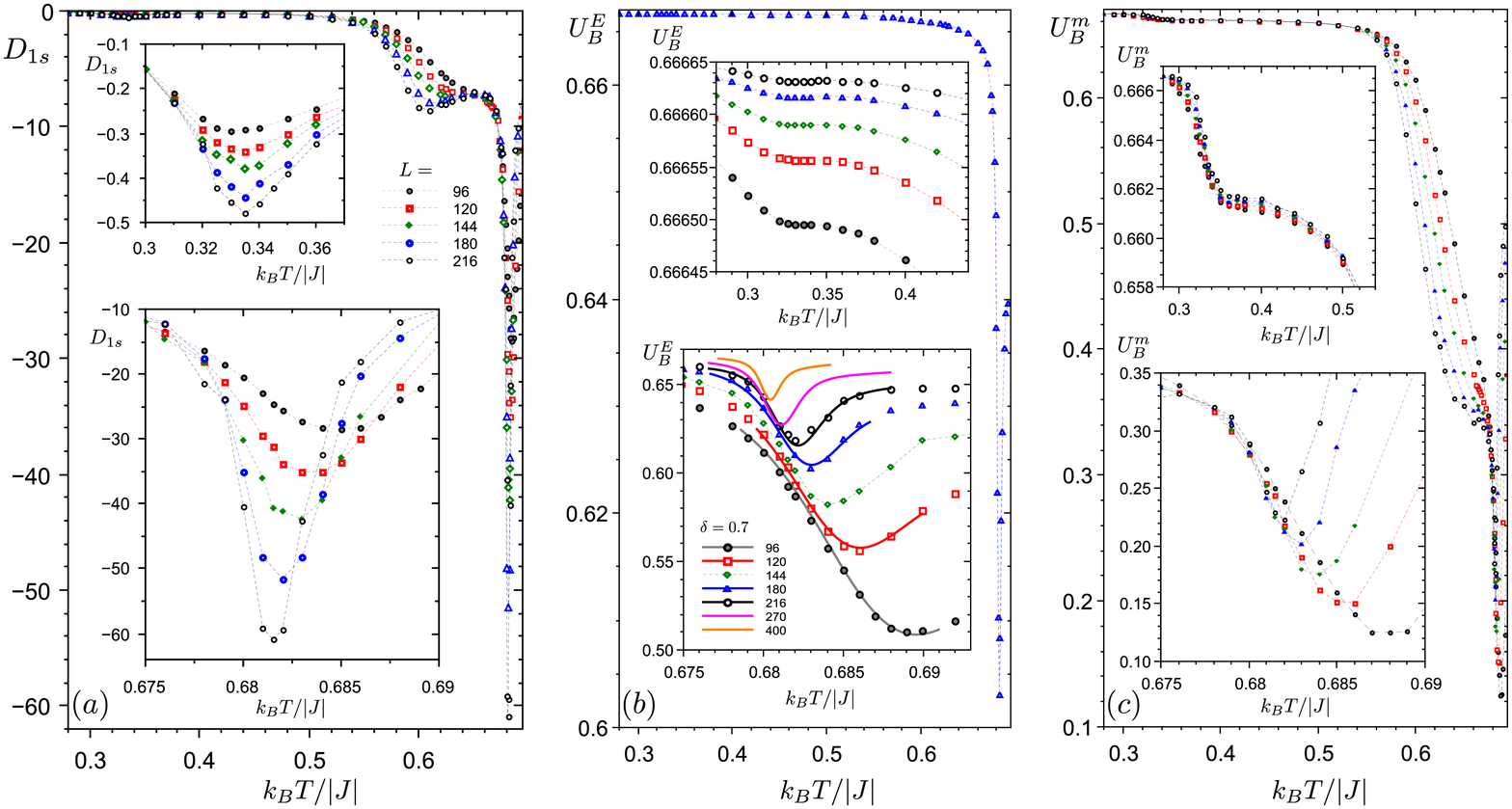}
\caption{(Color online) Temperature dependence of (a) parameter $D_{1s}$ and
Binder cumulants $U_B^E$ (b) and  $U_B^m$ (c) of the 3NN12 model at $\delta=0.7$
and different values of $L$. Magnified dependences close to transitions at $T_1$
and $T_c$ are shown in upper and lower insets, respectively. Symbols correspond
to thermal averaging results, dashed lines are guides to the eye. In lower inset
of (b) the results of histograms reweighting close to $T_c$ are shown by solid
lines.}  
\label{fig3.eps}
\end{figure} 

Temperature dependence of staggered magnetization (Fig. 2a) demonstrates that
3NN12 model has three phase transitions at $\delta=0.7$. In addition to two
transitions at $T_1$ and $T_2$ (which correspond to peaks of susceptibility in
Fig. 2b), the high temperature phase transition at $T_c$ is nicely visible as a
twist of $m_s(T)$ dependence at very low values of $m_s<0.05$. In temperature
dependence of susceptibility, the $T_c$ might be noticed as a small
higher-temperature shoulder of the peak at $T_2$. However, the transition at
$T_c$ corresponds to the main high-temperature peak in the $C_v(T)$ dependence
(Fig. 2c, d), and here the transition at $T_2$ is its hardly discernible
lower-temperature satellite (Fig. 2c). The transition at $T_1$ is very weakly $L$-dependent (Fig. 2e).

All three transitions are best manifested (see Fig. 3a) in temperature
dependences of parameters $D_{1s}$  and $D_{2s}$ (3) which combine the
contributions of energy and order parameter. The Binder cumulants of energy and
magnetization are shown in Figs. 3b and c. In both of them the transition at
$T_1$ is manifested as a smooth continuous step and the transition at $T_c$ - as
a deep minimum. The  transition at $T_2$ is not seen in $U_B^E$, but clearly
seen in $U_B^m$ in between the transitions at $T_1$ and $T_c$. 

\begin{figure}[] 
\includegraphics[width=\columnwidth]{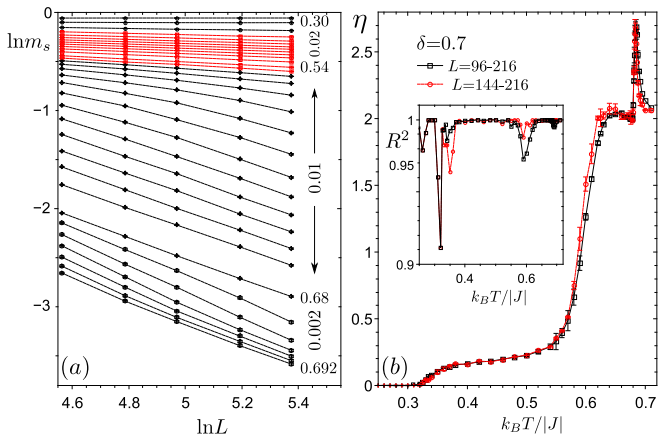} 
\caption{(Color online) (a) Log-log plot of $m_s$ vs $L$ for 3NN12 model at
$\delta=0.7$ in a temperature interval comprising the phase transition points at
$T_1$, $T_2$ and $T_c$. The BKT-type transition region is shown by red lines.
(b) Temperature dependence of parameter $\eta$ calculated for 5 (black curve)
and 3 largest (red curve) lattice sizes. Inset: temperature dependence of a
linear fit accuracy parameter $R^2$ for both cases.}
\label{fig4.eps}
\end{figure} 

Visually, the $m_s(T)$ dependence between the transition points at $T_1$ and
$T_2$  suggests similarity of this dependence to the 1NN model. Analysis of
log-log plots of magnetization vs $L$ (Fig.4a) corroborates the finding of the
1NN model that the transitions at $T_1$ and $T_2$ belong to the BKT-type phase
transitions. 
This is seen from the temperature dependence of the critical exponent of the
correlation function, $\eta$ (Fig. 4c), which for the BKT-type transitions
should correspond to the doubled slope of lines in Fig. 4a. In temperature range
between 0.35 and 0.55 the parameter $\eta$ clearly demonstrates a plateau. The
interval of $\eta$ values in the plateau roughly coincides with classical
predictions for the critical line of the BKT-type phase transitions~\cite{jose}.
We performed the FSS analysis using formula (4) to obtain more accurate values
of transition temperatures $T_1$ and $T_2$ and $\eta$. The results are shown in
Fig. 5a and b. The best fit was obtained for the values
$k_BT_1/|J|=0.35\pm0.01$, $\eta(T_1)=0.12\pm0.02$ and $k_BT_2/|J|=0.55\pm0.01$,
$\eta(T_2)=0.29\pm0.02$.
As for the 1NN model, the obtained value of $T_1$ is a bit higher than that at
the peak of $C_v$ and very similar to that at the peak of $\chi$, while $T_2$
lies lower than that obtained at the peak of $\chi$. It should be also noted,
that we expected some error in determination of the $T_2$ point, since in the
3NN12 model, differently from the 1NN model, the transition at $T_2$ is not from
the paramagnetic phase, but from the structure existing between $T_2$ and $T_c$
(see below). Still, as might be seen from Fig. 5b the scaling is quite
satisfactory.
 
\begin{figure}[] 
\includegraphics[width=\columnwidth]{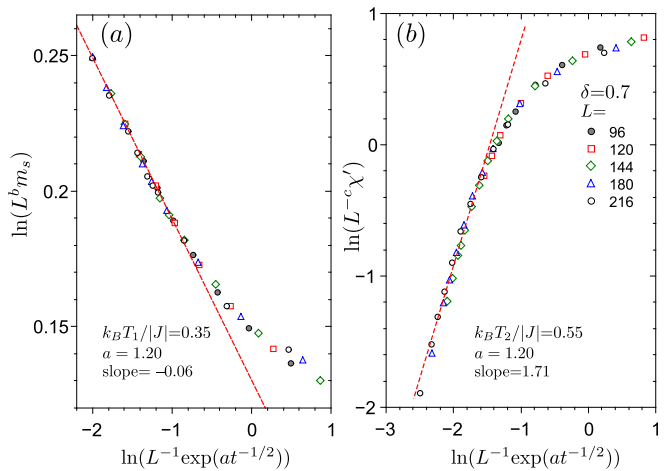}
\caption{(Color online) Finite-size scaling of (a) $m_s$ at $T_1$
($t=(T_1-T)/T_1$) and (b) $\chi'$ at $T_2$ ($t=(T-T_2)/T_2$) for 3NN12 model at
$\delta=0.7$ obtained using first and second scaling relations (4),
respectively.}  
\label{fig5.eps}
\end{figure} 

The $\eta$ interval of the BKT points is rather close to the one obtained in
similar models: the 1NN model (0.12-0.29)~\cite{zukovic}, the planar rotator
model with sixfold symmetry breaking fields (1/9-1/4)~\cite{jose}, 6-state AFM
clock model ((0.13-0.25)~\cite{surungan} and (0.1-0.275)~\cite{challa2}) and
TAFI model with 2NN ferromagnetic interactions (0.15-0.27)~\cite{landau}. 

The calculation of the Binder magnetic fourth-order cumulant $U_B^{m}$ also
demonstrated that transitions at $T_1$ and $T_2$ belong to universality class of
the BKT-type phase transitions. The $U_B^m(L)$ vs $U_B^m(L')$ plots  revealed
that basically $U_B^m(L)\rightarrow U_B^m(L')$ with increase of $L'<L$, and
consequently $\nu\rightarrow\infty$ in the formulae, $\partial
U_B^m(L')/\partial U_B^m(L) =(L'/L)^{1/\nu}$, as for the BKT-type phase
transition point.

It should be also noted that the $\eta(T)$  dependence (Fig. 4c) also ``feels''
the phase transition at $T_c$ demonstrating a sharp peak at the same value of
temperature where the extrema of $D_{1s}$, $D_{2s}$ and $C_v$ are obtained. This
is not unexpected: the point at $T_c$ is critical. The high value of $\eta$ at
$T_c$ makes it impossible to assign this transition to universality class of the
BKT-type of transitions, raising a challenging problem of its attribution. If
the phase transition would be of the second order, the order parameter at $T_c$
should scale as $\sim L^{-\beta/\nu}$. However, the value of $\beta/\nu$ is much
too large and inconsistent with the second-order phase transition, the
indication that the first-order phase transition might take place at $T_c$.

We noticed that this transition occurs at approximately same concentration of
particles as the stoichiometric concentration of the low-temperature LRO AFM
phase (see Fig. 6). Visual inspection of instant particle configuration reveals
marked increase (in comparison to the paramagnetic phase) of hexagons with side
length $2a$ and alternation of +1 and -1 variables on the vertices and 0 in the
center, i.e. hexagons typical to the low-$T$ phase of the 3NN12 model. These
domains of low-temperature phase exist in a very small interval of temperature
between $T_c$ and $T_2$. Decrease of temperature from $T_c$ leads to an increase
of concentration which results in population of centers of mentioned hexagons
and formation of a frustrated structure at $T_2$, the structure which further
continues up to the phase transition point at $T_1$. This increase of
concentration is rather abrupt in comparison to a smooth and continuous increase
of $c$ characteristic to the 1NN model (compare curves in Fig. 6). Thus, at
$T_c$ we 
obtain a strongly diluted phase with domains of the low-temperature AFM LRO
phase. It is known that dilution in frustrated systems leads to phase
transitions with non-classical critical exponents, broad two-maxima histograms
with high saddle point, ambiguous behaviour of interface energy and, in general,
makes the FSS analysis very complicated~\cite{janke}.

\begin{figure}[] 
\includegraphics[width=0.5\columnwidth]{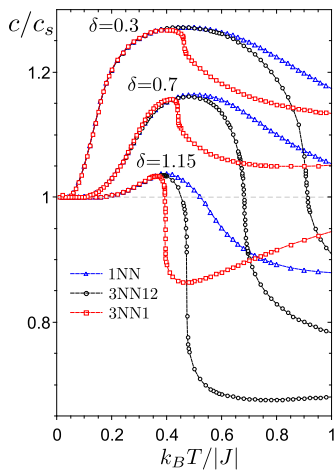}
\caption{(Color online) Temperature dependence of concentration for three models
at $L=120$: 1NN (blue triangles), 3NN12 (black squares) and 3NN1 (red squares)
at $\delta=0.3$, 0.7 and 1.15.}  
\label{fig6.eps}
\end{figure}

The energy histograms at the $T_c$ point and $\delta=0.7$ are shown in Fig. 7.
They are two-peaked and remain such up to the largest lattice size studied here,
$L=399$. We calculated interface tension,
$2\sigma=\ln(P_{\mathrm{max}}(L)/P_{\mathrm{min}}(L))/L$, and latent heat,
$\Delta E=|E_{+}-E_{-}|$, using these energy histograms. Here
$P_{\mathrm{max}}(L)$ and $P_{\mathrm{min}}(L)$ are probability density of
energy at maximum and saddle point, respectively, and $E_{+}(L)$ and $E_{-}(L))$
are the energies at right and left peaks of energy distribution at $T_c$. The
$2\sigma$ even up to $L=399$ depends on $L$ - thus, we are not sure if we have
reached the lattice sizes suitable for the finite size scaling, but $L=399$ was
the limit of our computer resources. The saddle point slightly decreases with
$L$ which would indicate in favor of the first order phase transition, though
the behavior is rather different from that of the typical first order phase
transition (which will be seen in region III). If to 
neglect the two smallest lattice sizes, the interface tension and latent heat, 
decrease with increase of $L$ as shown in insets to Fig. 7. The $\Delta E$, most
likely, tends to a finite value.

\begin{figure}[] 
\includegraphics*[width=\columnwidth]{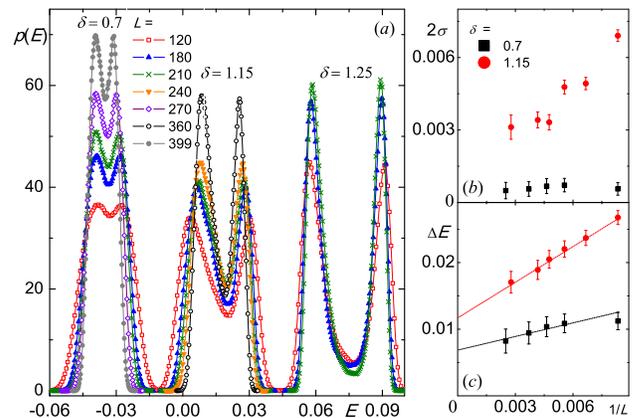}
\caption{(Color online) (a) Energy histograms of 3NN12 model at $\delta=0.7$,
1.15 and 1.25 (the latter is shifted along energy axis by 0.05). (b) and (c)
show $L$-dependence of interface tension and latent heat, respectively.}  
\label{fig7.eps}
\end{figure} 
 
The results of specific heat scaling close to $T_c$ are given in Fig. 8a. At $\delta=0.7$ the values of critical exponent ratios $\alpha/\nu=1.04\pm0.05$ and $1/\nu=1.64\pm0.05$ are obtained. The calculation of $\alpha/\nu$ from magnitude of $C_v$ peak at $T_c$ for lattice sizes $L=180-399$ yields the same result, $\alpha/\nu=1.04$ (see Fig. 8c). Calculation of  $1/\nu$ from minima of $D_{1s}$ and $D_{2s}$ at $T_c$ for lattice sizes $L=144-216$ gave us much smaller critical exponent of the correlation length, $1/\nu=1.0\pm0.05$. 

We also performed histograms and critical exponents calculation at another point of region I, $\delta=0.3$. The saddle point of these histograms is even higher than in the case $\delta=0.7$, correspondingly the interface tension is a bit
smaller. The latent heat is similar to that of $\delta=0.7$. The
scaling performed at $\delta=0.3$ gives the following critical exponents $\alpha/\nu=0.83\pm0.05 $ (the same value as from the fitting of $C_v(T_c)$, see Fig. 8c) and $1/\nu=1.5\pm0.05$. The critical exponents obtained here from scaling of    
$D_{1s}$ and $D_{2s}$ at $T_c$ give $1/\nu=1.0\pm0.05$, the same as for $\delta=0.7$.

\begin{figure}[] 
\includegraphics[width=\columnwidth]{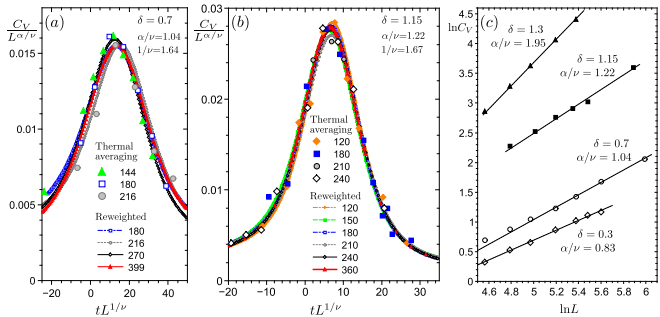} 
\caption{(Color online)  Finite-size scaling of specific heat at $T_c$ for the 3NN12 model: (a) $\delta=0.7$ and (b) 1.15. The results are fitted using formula $C_v-C_0\sim L^{\alpha/\nu}f(tL^{1/\nu})$, where $t=|T_c-T|/T_c$ and background is assumed to be $C_0=0$. Large symbols correspond to the results of thermal averaging, lines and small symbols - to results obtained close to $T_c$ by reweighted histogram method. (c) Log-log dependences of $C_v$ maximum vs $L$ at different values of $\delta$.}  
\label{fig8.eps}
\end{figure} 

Thus, in range I at $T_c$ we do not obtain standard values
$1/\nu=\alpha/\nu=d=2$ as for the first order phase transitions. In general, the
behavior of thermodynamic parameters at $T_c$ are much smoother in range I than
in ranges II and III. We assume that in the range I this transition is either a
weak first order phase transition, as often encountered in models with site or
bond dilution~\cite{janke}, or a second order phase transition with the latent
heat approaching to zero for such values of $L$ which exceed considerably our
computer resources (the value $1/\nu=1$ obtained from scaling of parameters $D_1$ and $D_2$ for $\delta=0.3$ and 0.7 is the same as for the Ising universality class). 
Thus, non-standard critical exponents obtained at $T_c$ in this range might be considered as a very crude approximation only. We do not exclude the possibility that the results would probably change for considerable increase of $L$. 

It should be noted though, that at $\delta=0.7$ the obtained set of critical exponents is rather close to the one obtained by Landau~\cite{landau} for tricritical region of the TAFI model with ferromagnetic 2NN interactions ($\alpha/\nu=1.02$ and $1/\nu=1.59$). In comparison, the theoretical predictions for the tricritical point of
the 3-state Potts model~\cite{nijs} are $\alpha/\nu=10/7=1.43$ and
$1/\nu=12/7=1.71$.

\subsection{3NN12 model at other values of $\delta$ (ranges II and III)}

The study of the 3NN12 model at other values of $\delta$ allowed to determine
the exact location of three ranges with different behavior of phase
transitions. We found that the ranges I ($T_c>T_2>T_1$), II ($T_c=T_2>T_1$) and
III ($T_c=T_2=T_1$) correspond to following ranges of $\delta$
values: 0 - 1.15, 1.15 - 1.3 and  1.3 - 1.5, respectively. The transitions at
$T_c$ in range III is clearly of the first order.
It is demonstrated by energy histograms for $\delta=1.3$ and 1.45 presented in Fig. 9 . The saddle point in this region is much lower than in region I and decreases with increase of $L$. For
$\delta\approx1.3-1.5$ this tendency only increases: at $\delta=1.45$ the peaks
are separated by a huge gap (no saddle at all). 

\begin{figure}[] 
\includegraphics*[width=\columnwidth]{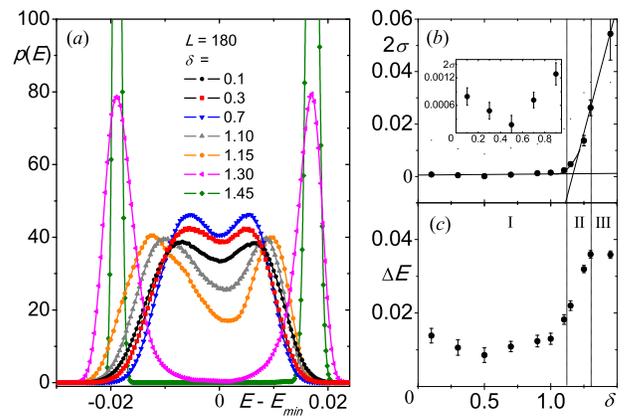}
\caption{(Color online) (a) Energy histograms at various values of $\delta$ at
$T_c$ for $L=180$ lattice. Corresponding  $\delta$-dependences of interface
tension and latent heat are given in (b) and (c), respectively.}  
\label{fig9.eps}
\end{figure} 

The same, just not so strong tendency to the first order phase transitions is
seen in range II. We have chosen the points $\delta=1.15$ and $\delta=1.25$ for
more thorough examination. The histograms at these points are presented in Fig.
7. The interface tension and latent heat are much higher than in range I. In
principle, with respect to the order of transition at $T_c$, this range is
intermediate between the ambivalent-order phase transition in range I and the
first-order phase transition in range III. If the transitions in range I turned
out to be of the second order, the range II would be the tricritical region. 

Different properties of phase transition at $T_c$ in ranges II and III, on one
hand, and I, on the other, might be seen analyzing the energy histograms (Fig.
9a) at fixed $L$ and various values of $\delta$, especially $\delta$-dependences
of interface tension $2\sigma$ and latent heat $\Delta E$ at $T_c$ (Fig. 9b and
c). It is seen that both these parameters clearly increase for higher values of
$\delta$. Here we can notice the separation of the system into three mentioned
ranges of behaviour: the range I featuring two-peaked histograms with high
saddle point, the range III demonstrating  typical first-order phase transition
and the intermediate range II. In Fig. 9b the intersection of two lines
corresponding to types of  behavior in ranges I and III is around 1.1-1.2 for
$L=180$. 
It should be noted here that even the behavior in $0<\delta<0.9$ region is not
so homogeneous as might be assumed from main Fig. 9b. The detailed inset in this
Fig. demonstrates that interface tension slightly increases when the limit of
the TAFI model, $\delta=0$, is approached and therefore has some minimum around
$\delta\approx0.5$, the minimum which survives also for other values of $L$. 

Rather similar result, demonstrating the division into several ranges of
behavior, is obtained
analyzing the magnitude of minimum related to $T_c$ of both Binder cumulants,
$U_B^m(T_c)$ and $U_B^E(T_c)$. They have two very different regions of behavior:
up to approximately $\delta=0.9$ the minimum of $U_B^m$ is around 0.3-0.1, but
drastically decreases for higher values of $\delta$. The minimum of $U_B^E$ is
rather close to the 2/3 limit up to $\delta=0.7$, but again start to rather
abruptly decrease at higher values of $\delta$.  

The results of our thermal averaging MC simulation in ranges II and III
demonstrate that the thermodynamic parameters close to $T_c$ either show thin
and high extrema ($C_v$ and $D_{1s}$, $D_{2s}$) or abruptness similar to jump
($m_s$ and average energy), see e.g. the behavior of normalized coverage at
$\delta=1.15$ in Fig. 6. Thus, these results just confirm the results obtained
by histogram calculations that the phase transitions in these two regions are of
the first order. 

In ranges II and III we also performed FSS analysis and determined the ratios of critical exponents. We obtained $\alpha/\nu=1.22$, 1.68 and 1.95 ($\pm 0.05$) and
$1/\nu=1.67$, 1.94 and 1.99 for $\delta=1.15$, 1.25 and 1.3 ($\pm 0.05$), respectively. Some results of this analysis are presented in Fig. 8b and c.
While the values of $\alpha/\nu$ and $1/\nu$ at limiting points of the range II, $\delta=1.25$ and 1.3, tend to the value 2  and are further stabilized at $d=2$ for $\delta>1.3$ (range III), the values at the other limiting point, $\delta=1.15$,
are closer to those of the range I (and point $\delta=0.7$, in particular).

\subsection{3NN1 model (ranges II and III)}

\begin{figure}[] 
\includegraphics[width=1.0\columnwidth]{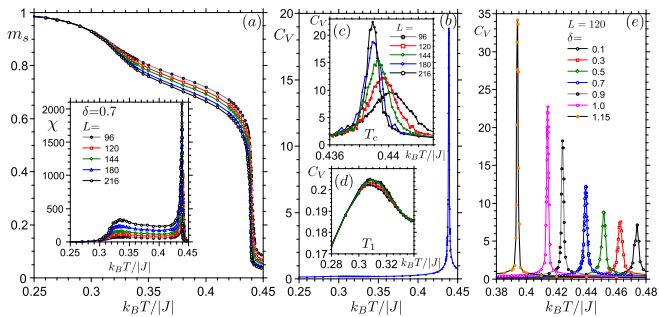} 
\caption{(Color online) Temperature dependence of (a) staggered magnetization,
(inset in (a)) susceptibility and (c, d) specific heat close to transition
points $T_2=T_c$ and $T_1$ of the 3NN1 model at $\delta=0.7$ and different
values of $L$. (b) The $C_v$ vs $T$ dependence at $\delta=0.7$ and $L=120$. (e)
The $C_v$ vs $T$ dependence at $L=120$ and different values of $\delta$.}  
\label{fig10.eps}
\end{figure} 
 
In the 3NN1 model the exclusion of only 1NN sites leaves more sites for
diffusion of particles, in comparison to the 1NN and 3NN12 models,
correspondingly increasing the configurational entropy of the system. This does
not affect the low-temperature transition point to the AFM LRO phase at $T_1$,
but decreases the high-temperature phase transition point from the paramagnetic
phase at $T_c$. As a result, $T_c$ falls into the critical line of the BKT-type
phase transition points and becomes inseparable from the high-temperature end of
this line, $T_2=T_c$, in interval of $\delta$ values, $0<\delta\lesssim1.2$
(region II). The first-order phase transition region III, where all transition
points coincide, $T_2=T_1=T_c$, is at $1.2\lesssim\delta<1.5$. Thus, at
$0<\delta\lesssim1.2$ the behavior of the 3NN1 model at $T_c$ is expected to be
similar to that of the 3NN12 model in region II ($1.15<\delta<1.3$). Thus,
the question arises if the point at $T_2=T_c$ is the higher-temperature end of
the BKT-type phase transitions, as in the 1NN model, or has the properties of
the first-order phase transition as in the region II of the 3NN12 model.

The $m_s(T)$ dependence at $\delta=0.7$ is given in Fig. 10a. There are no
qualitative changes in comparison to the 1NN and 3NN12 models at $T_1$, where we
can expect the low-temperature BKT-type phase transition. At $T_2=T_c$ the
$m_s(T)$ curve is weakly-dependent on $L$, which would make it a likely
candidate for the BKT-type transition. On the other hand, the dependence at this
point is very abrupt as in the case of the first-order phase transition. The
peak of susceptibility (inset in Fig. 10a) at $T_2=T_c$ is much higher than that of
the 3NN12 model at the $T_2<T_c$ , but comparable with the one obtained in
the 3NN12 model when $T_2=T_c>T_1$ at $1.15<\delta<1.3$. Specific heat $C_v$
demonstrates (Fig. 10b) a sharp peak at $T_c=T_2$ which depends on $L$ (Fig. 10c) and clearly increases with increase of $\delta$ (Fig. 10e). The $C_v$ also shows very small in comparison to the main peak and almost $L$-independent peak at $T_1$ (Fig. 10d).

\begin{figure}[] 
\includegraphics[width=\columnwidth]{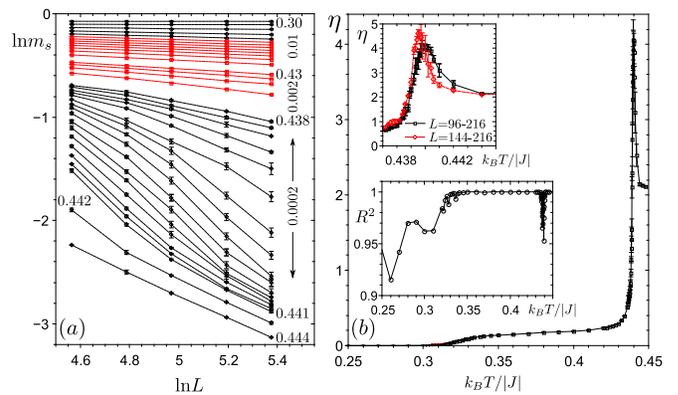} 
\caption{(Color online) (a) Log-log plot of $m_s$ vs $L$ for 3NN1 model at
$\delta=0.7$ in a temperature interval comprising the phase transition points at
$T_1$ and $T_2=T_c$. The BKT-type transition region is shown by red lines. (b)
Temperature dependence of parameter $\eta$ obtained from (a). Insets: (upper) 
$\eta(T)$ dependence close to $T_c$ peak for five (black curve) and three largest (red
curve) lattice sizes, respectively; (lower) temperature dependence of a linear
fit accuracy parameter $R^2$.}  
\label{fig11.eps}
\end{figure} 

Log-log plots of $m_s$ vs $T$ dependence (Fig. 11a) and, consequently, $\eta$ vs $T$ dependence in Fig. 11b clearly demonstrate the region of the BKT-type phase transitions
and $T_1$ as its low-temperature end ($\eta\sim 0.12$ at $T_1$). However, the high-temperature
end of the BKT-type transitions line shows a high peak at $T_2=T_c$ instead of
rounding which is characteristic to $T_2$ encountered in the 1NN and 3NN12
models.  Moreover, the FSS analysis of the phase transition point close to $T_1$  might
be performed (see Fig. 12a) using the first formula (4), and  the best fit
gives $T_1=0.34\pm0.01$ and $\eta(T_1)=0.12\pm0.02$. However, the FSS analysis using
second formula (4) close to $T_c=T_2$ is rather unsuccessful. The critical
behavior at the $T_c=T_2$ peak of $\eta(T)$ dependence is also inconsistent with any
reasonable $\beta/\nu$ values related to the second-order phase transitions. 
All these facts indicate that for the 3NN1 model the  transition at $T_1$ is the
BKT-type phase transition, but the high-temperature transition at $T_2=T_c$ - is
not. The latter conclusion is further confirmed by the histograms calculation.
The saddle point in two-peak energy histograms of Fig. 13a at $\delta=0.7$
rather systematically decreases with increase of $L$, supporting the idea of the
first-order phase transition at this point. 

\begin{figure}[] 
\includegraphics*[width=\columnwidth]{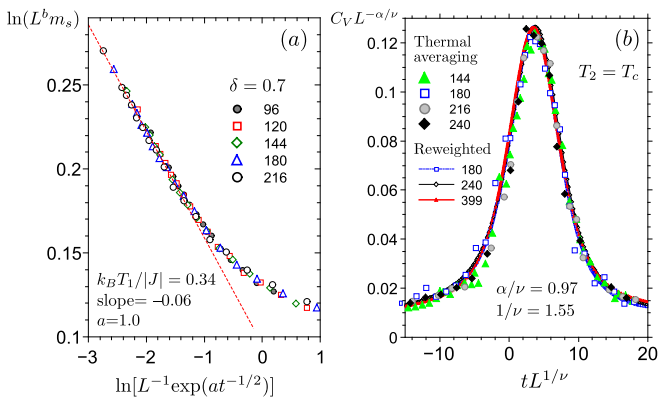}
\caption{(Color online) Finite-size scaling  of the 3NN1 model parameters at $\delta=0.7$: (a) $m_s$ at $T_1$ ($t=(T_1-T)/T_1$) and (b) $C_v$ at $T_c=T_2$ using scaling relation $C_v\sim L^{\alpha/\nu}f(tL^{1/\nu})$, where $t=|T_c-T|/T_c$.}
\label{fig12.eps}
\end{figure} 

\begin{figure}[] 
\includegraphics[width=\columnwidth]{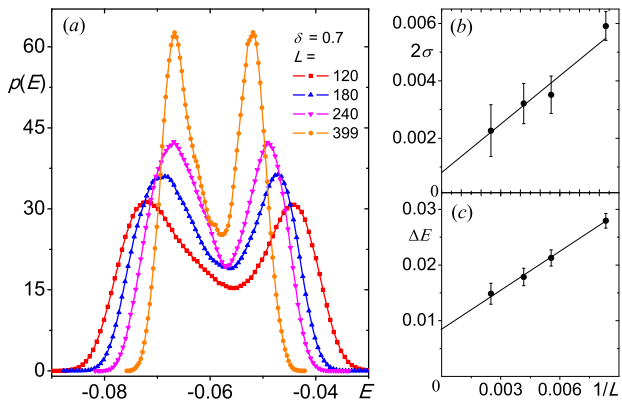} 
\caption{(Color online) (a) Energy histograms, (b) interface tension
$2\sigma=\frac{1}{L}\ln( P_{\mathrm{max}}/P_{\mathrm{min}})$ and (c) latent heat
of the 3NN1 model at $\delta=0.7$.}  
\label{fig13.eps}
\end{figure}

One more argument in favor of the first order phase transition at $T_2=T_c$
of the 3NN1 model in region II comes from analysis of the autocorrelation time of energy. As mentioned, the integrated autocorrelation time for the 3NN1 model at $\delta=0.7$ and $T_c=T_2$ is around one-two
orders of magnitude higher than for the 3NN12 model at $\delta=0.7$ and $T_c>T_2$ and is approximately $\tau\sim10^4-10^5$ for $L=120$ and close to the limit of our calculations, $\tau\sim10^6-10^7$, for $L=399$. Such difference
between the autocorrelation times for both models indicates much stronger first-order
nature of the transition at $T_2=T_c$ of the 3NN1 model.

Nevertheless, the extrema of $C_v$ and $D_{1s}, D_{2s}$ parameters in region II
at $T_2=T_c$ do not scale with $L^2$ as for the usual first-order phase transition. This is clearly seen extending the lattice sizes up to $L=399$. The $C_v$ scales with critical exponents $\alpha/\nu=0.97\pm0.05$ and $1/\nu=1.55\pm0.05$ (Fig. 12b).
The FSS of $D_{1s}$ and $D_{2s}$ parameters gives very similar result for
$1/\nu$. The same is true when $1/\nu$ is obtained from scaling of
$T_c(L)=T_c(\infty)+aL^{-1/\nu}$, when $T_c(\infty)$ is the one used for scaling
in Fig. 12b.

The first-order type of transition at $\delta>1.2$ (region III) is much more
explicit than at $\delta=0.7$. The critical exponents tend to 2 demonstrating
typical first-order behaviour at $\delta>1.2$. The extrema of $C_v$, $D_{1s}$
and $D_{2s}$ clearly scale as $\sim L^{2}$ when even smaller lattices are included  into our analysis.

\section{Discussion}

It is interesting to compare the phase transitions in 3NN12 and 3NN1 models with
those obtained in the 1NN model. This model has a critical line of the BKT-type
phase transitions (frustrated structure) in between temperature points $T_1$ and
$T_2$. At $T<T_1$ the AFM LRO phase is formed and at $T>T_2$ the phase is
paramagnetic~\cite{zukovic}. 

Both 3NN models also demonstrate the critical line of two BKT-type phase
transitions. As in the 1NN model, the temperature range between these two points
decreases with increase of a single-ion anisotropy parameter $\delta$, and 
these two BKT-type transitions flow into one first-order phase transition into
the AFM LRO phase at some $\delta_c$ (see phase diagram of all three models in
Fig. 14 a, b and d). The $\delta_c=1.47$~\cite{zukovic}, 1.3 and 1.2 for the 1NN,
3NN12 and 3NN1 models, respectively. The range at $\delta>\delta_c$ is called
the range III throughout this paper. 

As might be seen in Fig. 14, from $\delta=0$ and up to $\delta_c$ the 
phase transition temperature to the LRO AFM phase at $T_1$ (low temperature
end of the BKT-type phase transitions line) does not depend on the
model. 
 
\begin{figure*}[] 
\includegraphics*[width=1.7\columnwidth]{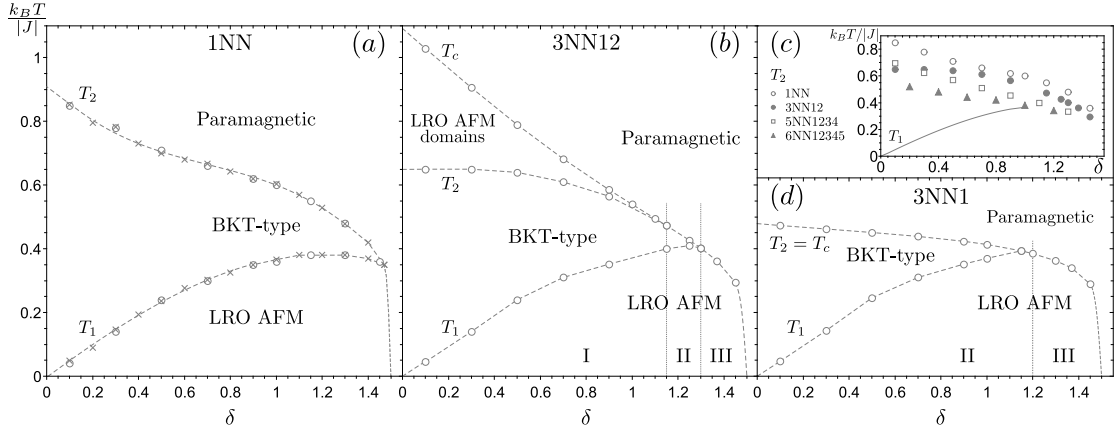}
\caption{(Color online) The phase diagrams of all three models: (a) 1NN ($L=48$), (b) 3NN12 ($L=180$) and  (d) 3NN1 ($L=180$). (c) The dependence of transition temperature $T_2$ on a chosen model with full exclusion (see text). The crosses in (a) are the
results of Ref.~\cite{zukovic}. Dashed lines are guides to the eye.} 
\label{fig14.eps}
\end{figure*} 

Correspondingly, the transition properties of the 1NN and 3NN12 models at lower
temperature are very similar in almost all interval of $\delta$ values. At
higher temperature and $\delta<\delta_c$ these two models have notable differences: 
where the 1NN model demonstrates higher temperature paramagnetic-to-BKT-type phase
transition
at $T_2$ and lower temperature transition at $T_1$, the 3NN12 model shows
three transitions: (i) phase transition at $T_c$ from paramagnetic phase to the
structure, which has the stoichiometry and separate domains of the LRO AFM
phase, (ii) higher-temperature transition to the BKT-type phase at $T_2$ and (iii) lower temperature transition at $T_1$. The difference in magnitude of $T_2$  occurs,
because the 3NN12 model has higher entropy, i. e. larger number of free sites
for hopping and higher probability of inhomogeneous distribution of particles
into domains. Therefore the frustrated phase in the 3NN12 model disorders at
lower temperature than in the 1NN model. 

The occurrence of phase transition at $T_c$ is related to the fact that
exclusions make the energy and other thermodynamic functions more abrupt at
higher temperature, i.e. exclusions create inhomogeneous distribution of
particles in disordered phase. Thus, income of particles is hindered and their
coverage $c$ is artificially maintained too small for that particular
temperature (in comparison to the 1NN model). Decrease of temperature 
enhances AFM correlations, and the ``normal'' coverage is recovered by sudden
increase of $c$. This might be seen e.g. in $c(T)$ dependences in Fig. 6 or
temperature dependences of internal energy which result in sharp peaks of
$C_v$ at $T_c$. Thus, the ``semi-ordered'' AFM phase has a chance to form in
the 3NN12 model at a bit higher temperature than the
frustrated phase occurs. The hump in $c(T)$ dependence marks the region of
frustrated phase between $T_1$ and $T_2$. In this temperature range the center
sites of the hexagons (which are formed by alternating  variables $\pm1$ on the
vertices) are partially filled. Thus, in the 3NN12 model the preconditions (relatively low temperature and stoichiometry corresponding to the AFM phase) allows for the AFM domains-phase to occur just before the hump. Higher concentration and correspondingly broader hump (lower values of $\delta$) shift the $T_c$ value to higher temperature,
while lower concentration ($\delta>1$) make the hump small and $T_c\rightarrow
T_2$. 

\begin{figure}[] 
\includegraphics[width=1.0\columnwidth]{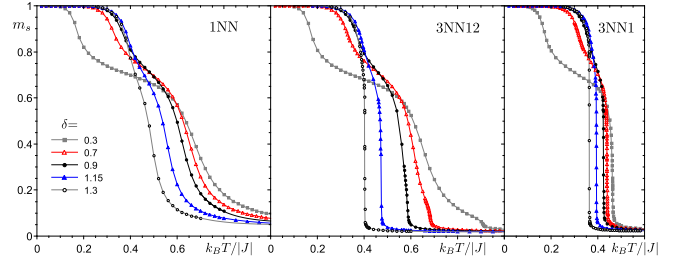} 
\caption{(Color online) Temperature dependence of staggered magnetization for
three models $\delta=0.3$, 0.7, 0.9, 1.15 and 1.3. The results were obtained
for lattice sizes $L=180$ (both 3NN models) and 48 (1NN model).}  
\label{fig15.eps}
\end{figure}

The entropy of the 3NN1 model (number of free sites for hopping) is even higher
than that of the 1NN or 3NN12 model. Therefore the temperature of the phase
transition from the paramagnetic phase for the 3NN1 model is the lowest of the
three models. Moreover, due to relative entropy increase, this phase transition
occurs at such a temperature which lies in temperature limits of the  line of
the the BKT-type critical points of the 1NN model. Therefore, contrary to the
1NN model, which has the line of critical BKT-type points between $T_1$ and
$T_2$, and the 3NN12 model, which (at least in part of $\delta$ interval)
demonstrates three phase transitions ($T_1$, $T_2$ and $T_c$), the 3NN1 model
shows reduced temperature interval of the BKT-type points line. The end point of
this line at $T_2$ coincides with $T_c$ for all values of $\delta$ and, as shown
by our analysis, most likely, does not belong to the BKT-type phase transitions.
Our analysis cannot distinguish, whether the two transitions merge into one point or they are separate transitions at extremely close temperatures with $T_2\lesssim T_c$, a  well-known situation for transitions to frustrated systems~\cite{noh,hasenbusch}. 
Note, that $T_c$ and $T_2$ for the 3NN12 model coincide only at narrow interval 
of $1.15<\delta<1.3$ (compare abrupt behaviour of $m_s(T)$ at this region of the
3NN12 model with that of the 3NN1 model in all interval of $\delta$ values, see
Fig. 15).  

Analyzing the obtained energy histograms at $T_c$ point, we found that the
histograms are two-peaked in all interval of $\delta$ values in both 3NN12 and
3NN1 models. However, the position of the saddle point in these histograms
clearly depends on $\delta$. The $\delta$-dependences of interface tension,
latent heat and Binder cumulants demonstrate that there are three ranges of
behavior of phase transitions. In range III the transition at $T_1=T_2=T_c$ is
clearly of the first order. This is evidenced by two-peaked histograms with a
saddle point which is either very deep or decreasing with increase of $L$. The
magnitudes of critical exponents $\alpha/\nu$ and $1/\nu$ are around 2. The
histograms in intermediate range II ($T_1<T_2=T_c$) have higher saddle point
than those in region III, but it also decreases with increase of $L$. This
allows to attribute the transition in this range to the first order, too.
However, in this range the finite size scaling at the $T_c$ point gives critical
exponents different from (though rather close to) 2. 

We could not present a definite answer about the type of the phase transition at
$T_c$ in region I of the 3NN12 model ($T_1<T_2<T_c$). The histograms are two-peaked here, but the
saddle point is rather high. Extrapolation of the values of interface tension
and latent heat at lattice sizes used in this paper shows that they approach
finite limits at $L\rightarrow\infty$. Rather controversial is the
$L$-dependence of the order parameter which indicates that interpretation of the
phase transition at $T_c$ in terms of critical exponents might be inconsistent.
This would allow to attribute this transition to a ``weak'' first-order phase
transitions observed in some diluted and frustrated systems~\cite{janke}.
Nevertheless, we cannot completely rule out the possibility of a second-order
phase transition manifesting itself as vanishing of the latent heat at much
larger lattices than used here.

It should be noted that the 3NN12 model is a unique model in which the splitting of
the  higher temperature phase transition into two transitions, at $T_c$ and
$T_2$, occurs. In addition to this model, two other models with full exclusions up to interaction
distance were also studied: the 5NN model with exclusions up to 4NN
(5NN1234) and 6NN model with exclusions up to 5NN (6NN12345). All models except 3NN12
demonstrate one higher temperature phase transition at $T_2=T_c$.
The transition temperature at $T_2$ and the parameter $\delta_c$ in all models
with full exclusions for entropic reasons gradually decreases with increase of
interaction distance of the model (see Fig. 14c). For $\delta<\delta_c$ the transition at $T_2$ is the higher-temperature BKT-type phase transition for the 1NN (no $T_c$) and 3NN12 ($T_2<T_c$) models, and of the first order for the 5NN1234 ($T_2=T_c$) and 6NN12345 ($T_2=T_c$) models.

Similar phase diagram as of the 3NN12 model might be observed in other frustrated systems. The phase transitions from paramagnetic to fully frustrated (FF) phase in square $\phi^4$ FFXY model~\cite{hasenbusch} proceed either through (i) Ising and BKT phase transitions sequence with very close transition temperatures (at small values of parameter $D$ similar to our $\delta$), (ii) tricritical region (intermediate values of $D$) featuring histograms with high saddle point or (iii) first order phase transitions (high values of $D$).

Here we studied strongly diluted lattices with small concentration of particles
(spins). Such studies require huge computer resources and therefore might leave
some questions not completely answered. However, the main tendencies are quite
clear: exclusions do not affect the low temperature phase transition; they make
the high temperature phase transition more abrupt; rescaling of lattice
stimulate the entropic effects and decreases the high temperature phase
transition temperature; in a case of the 3NN12 model the formation of domains of low-temperature structure might reveal itself as a phase transition at high temperature.

\section{Acknowledgements}

We are grateful to Wolfhard Janke for reading of the manuscript and valuable
discussions. A. Ibenskas acknowledges funding support by European Union
Structural Funds project “Postdoctoral Fellowship Implementation in Lithuania”
(VP1-3.1-\v{S}MM-01-V-02-004).




\begin{thebibliography}{}

\bibitem{bartels}
L. Bartels, Nature Chemistry {\bf 2}, 87 (2010).

\bibitem{barth}
J. V. Barth, Annu. Rev. Phys. Chem. {\bf 58}, 375 (2007). 

\bibitem{dmitriev}
A. Dmitriev, N. Lin, J. Weckesser, J.V. Barth, and K. Kern, J. Phys. Chem. B
{\bf 106}, 6907 (2002).

\bibitem{li1}
Z. Li, B. Han, L.J. Wan, and Th. Wandlowski, Langmuir {\bf 21}, 6915 (2005).

\bibitem{ye}
Y. C. Ye,  W. Sun, Y. F. Wang, X. Shao, X. G. Xu, F. Cheng, J. L. Li, and K. Wu,
J. Phys. Chem. C {\bf 111},  10138 (2007).

\bibitem{griessl}
S. Griessl, M. Lackinger, M. Edelwirth, M. Hietschold, and W.M. Heckl, Single
Mol. {\bf 3}, 25 (2002).

\bibitem{lackinger1}
M. Lackinger, S. Griessl, W.M. Heckl, M. Hietschold, and G.W. Flynn, Langmuir
{\bf 21}, 4984 (2005).

\bibitem{nath}
K.G. Nath, O. Ivasenko, J.M. MacLeod, J.A. Miwa, J.D. Wuest, A. Nanci, D.F.
Perepichka, and F. Rosei, J. Phys. Chem. C {\bf 111}, 16996 (2007).

\bibitem{kampschulte}
L. Kampschulte, T. L. Werblowsky, R. S. K. Kishore, M. Schmittel, W. M. Heckl,
and M. Lackinger, J. Am. Chem. Soc. {\bf 130}, 8502 (2008).

\bibitem{gutzler}
R. Gutzler, T. Sirtl, J. F. Dienstmaier, K. Mahata, V. M. Heckl, M. Schmittel,
and M. Lackinger, J. Am. Chem. Soc. {\bf 132}, 5084 (2010).

\bibitem{theobald}
J. A. Theobald, N. S. Oxtoby, M. Phillips, N. R. Champness, P. H. Beton, Nature
{\bf 424}, 424 (2003).

\bibitem{weber}
U. K. Weber, V. M. Burlakov, L. M. A. Perdigao, R. H. J. Fawcett, P. H. Beton,
N. R. Champness, J. H. Jefferson, G. A. D. Briggs, and D. G. Pettifor, Phys.
Rev. Lett. {\bf 100}, 156101 (2008).

\bibitem{silly}
F. Silly, U. K. Weber, A. Q. Shaw, V. M. Burlakov, M. R. Castell, G. A. D.
Briggs, and D. G. Pettifor,  Phys. Rev. B, {\bf 77}, 201408 (2008).

\bibitem{li2}
Y. Li, Z. Ma, G. Qi, Y. Yang, Q. Zeng, X. Fan, C. Wang, and W. Huang, J. Phys.
Chem. C {\bf 112}, 8649 (2008).

\bibitem{pawin}
G. Pawin, K. L. Wong, K.-Y. Kwon, and L. Bartels, Science, {\bf 313}, 961
(2006).

\bibitem{misiunas}
T. Misi\={u}nas and E. E. Tornau, J. Phys. Chem. B {\bf 116}, 2472 (2012).

\bibitem{bell}
G. M. Bell and D. A. Lavis, J. Phys. A: Gen. Phys. {\bf 3}, 568 (1970).

\bibitem{fiore}
C. E. Fiore, M. M. Szortyka, M. C. Barbosa, and V. B. Henriques, J. Chem. Phys. {\bf 131}, 164506 (2009).

\bibitem{blumecapel}
M. Blume, Phys. Rev. {\bf 141}, 517 (1966); H. W. Capel, Physica (Utr.) {\bf 32}, 966 (1966).

\bibitem{mahan}
G. D. Mahan and S. M. Girvin, Phys. Rev. B {\bf 17}, 4411 (1978).

\bibitem{beg}
M. Blume,  V. J. Emery, and R. B. Griffiths, Phys. Rev. A, {\bf 4}, 1071 (1971);
J. Sivardiere and J. Lajzerowicz, Phys. Rev. A {\bf 11}, 2090, 1975.

\bibitem{young}
A. P. Young and D. A. Lavis, J. Phys. A: Gen. Phys. {\bf 12}, 229 (1979).

\bibitem{barbosa}
M. A. A. Barbosa and V. B. Henriques, Phys. Reb. E {\bf 77} 051204 (2008).

\bibitem{wannier}
R. M. F. Houtappel, Physica (Utr.) {\bf 16}, 425 (1950); G. H. Wannier, Phys.
Rev. {\bf 79}, 357 (1950); R. J. Baxter, J. Phys. A {\bf 13}, L61 (1980).

\bibitem{ibenskas}
A. Ibenskas and E. E. Tornau, Phys. Rev. E {\bf 86}, 051118 (2012). 

\bibitem{petrauskas}
V. Petrauskas, S. Lapinskas, and  E. E. Tornau, J. Chem. Phys. {\bf 120}, 11815
(2004).

\bibitem{fortuna}
S. Fortuna, D. L. Cheung, and A. Troisi, J. Phys. Chem. B {\bf 114}, 1849
(2010).

\bibitem{simenas1}
M. \v Sim\.enas, A. Ibenskas, and E. E. Tornau, Phase Transitions {\bf 86}, 866
(2013).

\bibitem{simenas2}
M. \v Sim\.enas and E. E. Tornau, J. Chem. Phys. {\bf 139}, 154711 (2013).

\bibitem{schick}
M. Schick, J. S. Walker, and M. Wortis, Phys. Rev. B {\bf 16}, 2205 (1977); N.
Berker, S. Ostlund, and F. A. Putnam, Phys. Rev. B {\bf 17}, 3650 (1978).

\bibitem{salinas}
W. F. Wreszinski and S. R. A. Salinas, Disorder and competition in soluble
lattice models, Series on advances in statistical mechanics, vol. 9, World
Scientific, 1993.

\bibitem{miyashita}
S. Miyashita, Proc. Jpn. Acad., Ser. B {\bf 86}, 643 (2010); S. Miyashita, H.
Kitatani, and Y. Kanada, J. Phys. Soc. Jpn {\bf 60}, 1523 (1991).

\bibitem{cardy}
J. L. Cardy, J. Phys. A: Math. Gen. {\bf 13}, 1507 (1980) .

\bibitem{noh}
J. D. Noh, H. Rieger, M. Enderle, and K. Knorr, Phys. Rev. E {\bf 66}, 026111
(2002).

\bibitem{surungan}
T. Surungan, Y. Okabe and Y. Tomita, J. Phys. A: Math. Gen. {\bf 37}, 4219
(2004).

\bibitem{zukovic}
M. \v{Z}ukovi\v{c} and A. Bob\'ak, Phys. Rev. E {\bf 87}, 032121 (2013).

\bibitem{coddington}
P. D. Coddington and L. Han, Phys. Rev. B {\bf 61}, 2635 (1994).

\bibitem{reweight}
A. M. Ferrenberg and A. M. Swendsen, Phys. Rev. Lett. {\bf 61}, 2635 (1988);
{\bf 63}, 1658 (1989).

\bibitem{ferrenberg}
A. M. Ferrenberg and D. Landau, Phys. Rev. B {\bf 44}, 5081 (1991).

\bibitem{challa} 
M. S. S. Challa, D. P. Landau, and K. Binder, Phys. Rev. B {\bf 34}, 1841
(1986).

\bibitem{KT}
J.M. Kosterlitz and D.J. Thouless, J. Phys. C: Solid State Phys. {\bf 6}, 1181
(1973).

\bibitem{challa2}
M. S. S. Challa and D. P. Landau, Phys. Rev. B {\bf 33}, 437 (1986).

\bibitem{jose}
J.V. Jos\'e, L.P. Kadanoff, S. Kirkpatrick, and D.R. Nelson, Phys. Rev. B {\bf
16}, 1217 (1977).

\bibitem{landau}
D. P. Landau, Phys. Rev. B {\bf 27}, 5604 (1983).

\bibitem{janke}
C. Chatelain, P.-E. Berche, B. Berche, and W. Janke, Comp. Phys. Comm. {\bf
147}, 431 (2002);
S. Jin, A. Sen, W. Guo, and A. W. Sandvik, Phys. Rev. B {\bf 87}, 144406 (2013);
A. Kalz and A. Honecker, Phys. Rev. B {\bf 86}, 134410 (2012).

\bibitem{nijs}
M. P. M. den Nijs, J. Phys. A {\bf 12}, 1857 (1979); B. Nienhuis, A. N. Berker,
E. K. Riedel, and M. Schick,
Phys. Rev. Lett. {\bf 43}, 737 (1979); B. Nienhuis, J. Phys. A {\bf 15}, 199
(1982).

\bibitem{hasenbusch}
M. Hasenbusch, A. Pelissetto, and E. Vicari, J. Stat. Mech.  {\bf P12002}, 1 (2005);
Phys. Rev. B {\bf 72}, 184502 (2005).

\end{thebibliography}
\end{document}